\documentclass[aps,pra,twocolumn,superscriptaddress,10pt,article,nofootinbib,showpacs]{revtex4-2}
\usepackage[utf8]{inputenc}

\usepackage{natbib}
\usepackage{graphicx}
\usepackage{latexsym}
\usepackage{amssymb}
\usepackage{amsmath}
\usepackage{amsfonts}
\usepackage{gensymb}
\usepackage{bm}
\usepackage{hyperref}
\usepackage{braket}
\usepackage{physics}
\usepackage{bbold}
\usepackage{slashed}
\usepackage[dvipsnames]{xcolor}
\usepackage[normalem]{ulem}
\usepackage{siunitx}
\usepackage[version=4]{mhchem}

\DeclareRobustCommand{\rchi}{{\mathpalette\irchi\relax}}
\newcommand{\irchi}[2]{\raisebox{\depth}{$#1\chi$}}

\newcommand{\ii}{\mathrm{i}\,}
\newcommand{\vect}[1]{{\bm{#1}}}
\newcommand{\bq}{\vect{q}}
\newcommand{\bk}{\vect{k}}

\newcommand{\ts}{\mathrm{ts}}
\newcommand{\MIM}{\mathrm{MIM}}

\hypersetup{
  pdfauthor = {},
  pdftitle = {Wang et al. - 2023 - }
}

\newcommand{\supplementarysection}{%
  \setcounter{figure}{0}
  \let\oldthefigure\thefigure
  \renewcommand{\thefigure}{S\oldthefigure}
  \setcounter{section}{0}
  \let\oldthesection\thesection
  \renewcommand{\thesection}{S\oldthesection}
  \setcounter{equation}{0}
  \let\oldtheequation\theequation
  \renewcommand{\theequation}{S\oldtheequation}
}

\begin{document}

\title{Theory of the microwave impedance microscopy of Chern insulators}

\author{Taige Wang}
\affiliation{Department of Physics, University of California, Berkeley, CA 94720, USA \looseness=-1}
\affiliation{Material Science Division, Lawrence Berkeley National Laboratory, Berkeley, CA 94720, USA \looseness=-1}
\affiliation{Department of Physics, University of California, San Diego, CA 92093, USA \looseness=-1}

\author{Chen Wu}
\affiliation{Department of Physics, University of California, San Diego, CA 92093, USA \looseness=-1}

\author{Masataka Mogi}
\affiliation{Department of Physics, Massachusetts Institute of Technology, Cambridge, MA 02139, USA \looseness=-1}

\author{Minoru Kawamura}
\affiliation{Strong Correlation Physics Research Group, RIKEN Center for Emergent Matter Science, Saitama 351-0198, Japan \looseness=-1}

\author{Yoshinori Tokura}
\affiliation{Strong Correlation Physics Research Group, RIKEN Center for Emergent Matter Science, Saitama 351-0198, Japan \looseness=-1}
\affiliation{Department of Applied Physics and Tokyo College, University of Tokyo, Tokyo 113-8654, Japan \looseness=-1}

\author{Zhi-Xun Shen}
\affiliation{Department of Applied Physics, Stanford University, Stanford, CA 94305, USA \looseness=-1}
\affiliation{Geballe Laboratory for Advanced Materials, Stanford University, Stanford, CA 94305, USA \looseness=-1}
\affiliation{Stanford Institute for Materials and Energy Sciences, Stanford Linear Accelerator Center National Accelerator Laboratory, Menlo Park, CA 94025, USA \looseness=-1}

\author{Yi--Zhuang You}
\affiliation{Department of Physics, University of California, San Diego, CA 92093, USA \looseness=-1}

\author{Monica T. Allen}
\affiliation{Department of Physics, University of California, San Diego, CA 92093, USA \looseness=-1}

\date{\today}

\begin{abstract} 

Microwave impedance microscopy (MIM) has been utilized to directly visualize topological edge states in many quantum materials, from quantum Hall systems to topological insulators, across the GHz regime. While the microwave response for conventional metals and insulators can be accurately quantified using simple lumped-element circuits, the applicability of these classical models to more exotic quantum systems remains limited. In this work, we present a general theoretical framework of the MIM response of arbitrary quantum materials within linear response theory. As a special case, we model the microwave response of topological edge states in a Chern insulator and predict an enhanced MIM response at the crystal boundaries due to collective edge magnetoplasmon (EMP) excitations. The resonance frequency of these plasmonic modes should depend quantitatively on the topological invariant of the Chern insulator state and on the sample’s circumference, which highlights their non-local, topological nature. To benchmark our analytical predictions, we experimentally probe the MIM response of quantum anomalous Hall edge states in a \ce{Cr}-doped \ce{(Bi,Sb)2Te3} topological insulator and perform numerical simulations using a classical formulation of the EMP modes based on this realistic tip-sample geometry, both of which yield results consistent with our theoretical picture. We also show how the technique of MIM can be used to quantitatively extract the topological invariant of a Chern insulator, disentangle the signatures of topological versus trivial edge states, and shed light on the microscopic nature of dissipation along the crystal boundaries.

\end{abstract}


\maketitle

\section{Introduction}

Chern insulators host chiral one-dimensional edge states and exhibit a quantized Hall conductance due to a non-trivial topological band structure. The experimental signatures of Chern insulators have been observed in a variety of material systems, from the quantum Hall family to magnetic topological insulators, and more recently moiré materials \cite{KlitzingReview,AndreaReview,ShouChengQAH,QiKunQAH,YuanboDis,DGGmoire,Andreamoire,tMBG,Fengmoire,tTMD}. A key feature of Chern insulators is the presence of topological edge modes, which are electronic states that propagate unidirectionally along the edges of the material without backscattering. 
While edge states in Chern insulators have been investigated extensively using electronic transport techniques\cite{FoxonQPC,HarrisQPC,ButtikerQPC,MahaluQPC}, these methods lack the spatial resolution to probe the detailed structure and the degree of localization of these modes.

To address this limitation, several imaging techniques have been used to directly visualize topological edge modes, including scanning tunneling microscopy (STM) \cite{STM1,STM2,STM3,STM4,STM5}, superconducting quantum interference device (SQUID) microscopy \cite{PhysRevLett.113.026804, InAs_trivial, HgTe_SQUID}, as well as some interferometry techniques \cite{Graphene_edges}. These experiments have detected current flow and an enhancement of the density of states at the boundaries of topological insulators, which have been interpreted as evidence for one-dimensional topological edge modes. However, this interpretation can be complicated by the presence of trivial electronic states at the physical boundaries arising from impurities, dangling bonds, and  band bending \cite{Ma2015,EdgeDefect1,EdgeDefect2}.

In recent years, a near-field imaging technique called scanning microwave impedance microscopy (MIM) has shown great potential for spatially-resolved detection of topological boundary modes. 
Experiments have reported an enhanced microwave response at the edges of two-dimensional topological insulators and quantum Hall systems \cite{ShenspinHall, ShenHall, Monica,JianQAH,Ma2015,PhysRevLett.117.186601}, but the observed behavior cannot be easily explained by a simple conductance increase close to the edge using classical lumped-circuit models. Furthermore, the observed width of quantum Hall edge states, as measured by MIM, is an order of magnitude larger than that measured by transport and STM, significantly exceeding the magnetic length \cite{WangLocalization,ChangLocalization,PhysRevB.86.195417,STM1,STM2,STM3,STM4,STM5,Kim2021,PhysRevB.107.115426}.
This motivates our development of a theoretical foundation to compute the microwave response of quantum materials that cannot simply be characterized by a scalar conductivity value.

In this paper, we develop a general theoretical framework that quantifies the MIM response of a quantum material within linear response theory. 
Upon applying this theory to the special case of quantum anomalous Hall (QAH) insulators, we predict that an enhanced MIM response at the edge should arise from collective edge magnetoplasmon (EMP) modes that circulate along the sample boundary\cite{MacDonaldEMP,EMPExpt1,EMPExpt2,DGG,VolkovEMP,Justin,Liang,Chetan}. 
The resonance frequency of these plasmonic modes should depend quantitatively on the topological invariant of the Chern insulator state and on the length of the sample's perimeter \cite{DGG,MacDonaldEMP,EMPExpt1,EMPExpt2}. 
This non-trivial frequency dependence can unambiguously relate the enhanced edge signal observed with MIM to the one-dimensional topological edge modes that propagate around the entire sample perimeter, whereas topologically trivial edge effects are expected to be featureless in the frequency domain. 

To check the validity of this analytical model, we experimentally measured the real-space MIM response of the QAH edge modes in a \ce{Cr}-doped \ce{(Bi,Sb)2Te3} magnetic topological insulator at various microwave frequencies and also conducted numerical simulations based on this experimental tip-sample setup, both of which yielded results consistent with the theoretical framework proposed here.


\section{Theory of the MIM response of quantum materials} \label{sec:theory}

\begin{figure}
    \centering
    \includegraphics[width = 0.48\textwidth]{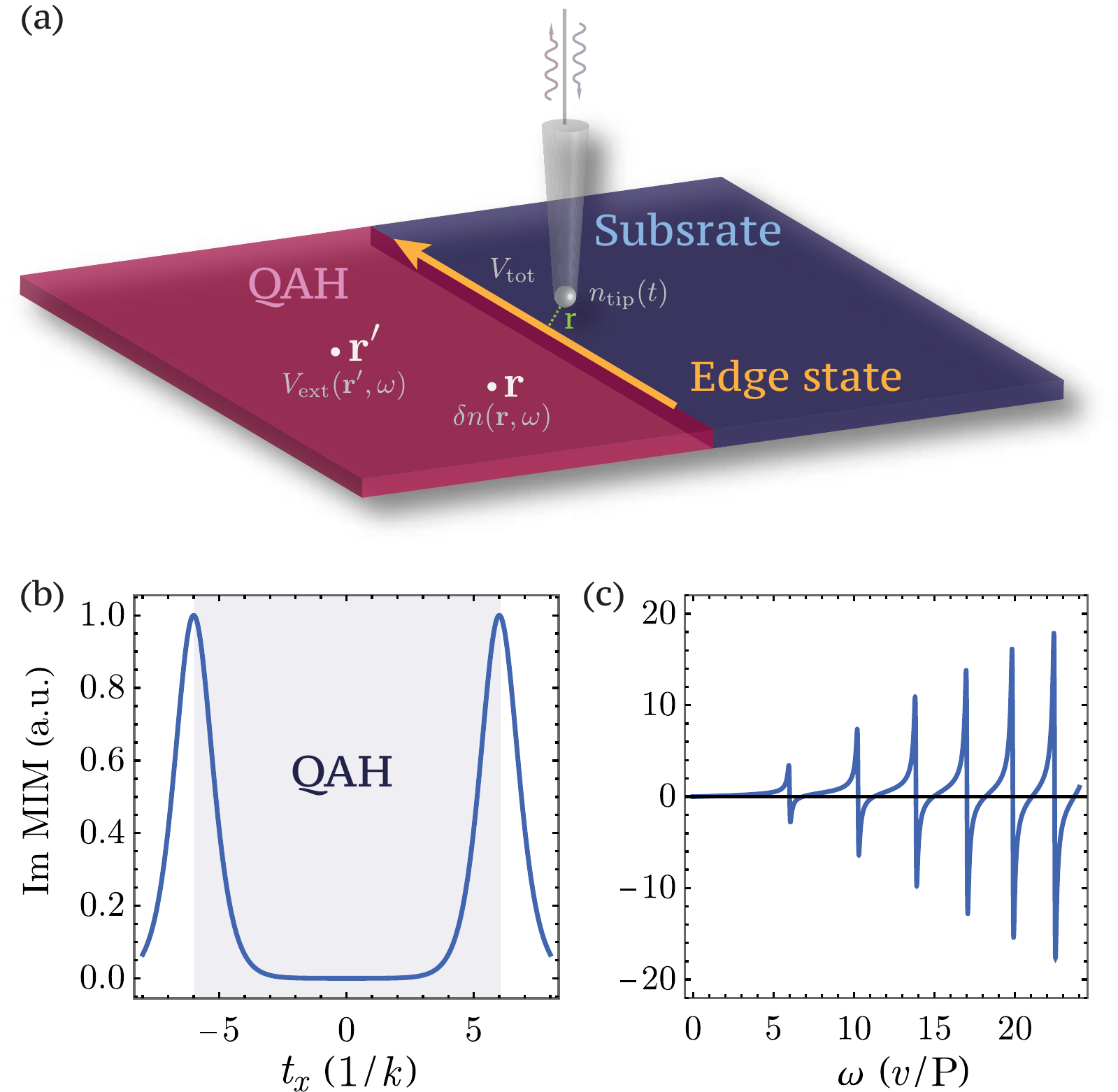}
    \caption{\textbf{Theoretical MIM response of topological edge modes in a quantum anomalous Hall insulator.} \textbf{(a)} Schematic illustration of the measurement setup, in which the MIM probe is scanned across the QAH insulator edge. The probe is driven by an AC voltage at microwave frequencies and then the displacement current is measured. The MIM signal can be computed by convolving with the correlation function $\rchi(\vect{r},\vect{r}')$ inside the sample. \textbf{(b-c)} The imaginary part of the MIM response of a QAH insulator according to Eqn.~(\ref{eq:MIMQAH}). In (b), the tip is scanned across the sample at fixed frequency $\omega$ and the sample lies within the shaded region (between $x = \pm 6/k_{\omega}$). Here $t_x$ is the tip location in units of $1/{k_{\omega}}$ with $k_{\omega}$ being the EMP momentum defined by $\omega \sim \omega_{\mathrm{EMP}}(k_{\omega})$. In (c), the tip is situated over the sample edge ($r = d$) while the frequency $\omega$ is swept. We used $d = 0.005 P$, $l = 0.2 P$, and $e^2/((2\pi)^2 \epsilon_0) = 0.2 \hbar v$, where $v$ is the edge velocity, and $P$ is the sample perimeter. In both cases, parameters are chosen to be relevant to the numerical simulation.}
    \label{fig:MIMTheory}
\end{figure}

MIM characterizes a material's electronic response to microwave frequency electromagnetic fields confined to a small spatial region around a sharp metallic probe. In practice, a microwave excitation is coupled to an AFM tip in close proximity to the sample, and the real and imaginary parts of the reflected signals are measured using GHz lock-in detection techniques \cite{Eric, MIMReview, SM}.

We first review the basics of the MIM measurement setup and then present a general model of the MIM response of quantum materials within linear response theory. As shown in Fig.~\ref{fig:MIMTheory}, the MIM probe, driven by an AC voltage at microwave frequency, is brought near the sample surface. Unlike scanning tunneling microscopy (STM), the probe is only capacitively coupled to the sample. MIM measures the displacement current exchanged between the probe and the sample. This displacement current $I$ can be formally written as $I = Y_{\ts} V$, where $V$ is the driven voltage and $Y_{\ts}$ is the complex tip-sample admittance \cite{Eric}.

In practice, an impedance matching network is always necessary to maximize the sensitivity of the admittance measurement. Since $Y_{\ts}$ is much smaller than the self-admittance of the MIM probe $Y_{t}$, the MIM signals are linearly proportional to the change in $Y_{\ts}$. Explicitly \cite{Eric},
\begin{equation} \label{eq:ab}
\Re \MIM + \ii \Im \MIM \approx a\left(\Re Y_{\mathrm{ts}}+\ii \Im Y_{\mathrm{ts}}\right)+b
\end{equation}
where $a$ is a real constant and $b$ is a complex constant. In the following section, we will show how to obtain $Y_{\ts}$ in terms of the density response function of the system.

\subsection{General framework}

The most commonly used model of $Y_{\ts}$ is the lumped-element model, which treats the sample as a resistor and a capacitor in parallel, then coupled capacitively to the tip \cite{SM}. This model can accurately predict the MIM response of conventional metals, dielectrics, and certain two-dimensional materials \cite{MIMReview}. However, a more general theoretical framework is required to properly quantify the microwave response of more complex quantum materials, such as topological states of matter, that cannot be simply described by a pair of scalar conductivity and permittivity values. For example, in the case of QAH insulators, the model must accommodate one-dimensional gapless plasmon modes at the sample edge, which is outside the scope of the traditional lumped element picture.

We start by approximating the tip apex by a single point at $\vect{r}_t$ and ignoring the rest of the tip (see Fig.~\ref{fig:MIMTheory}). This approximation is well justified for shielded tips and qualitative sound for other type of metallic probes. Then the external potential generated by the tip is given by $V_{\mathrm{ext}}(\vect{r},t) = G(\vect{r}, \vect{r}_t) Q_t(t)$, where $Q_t(t)$ is the external charge at the apex of the tip, and $G(\vect{r}, \vect{r}')$ is the Coulomb interaction inside the dielectric environment, which can either be calculated numerically, or be approximated by the vacuum value $G(\vect{r}, \vect{r}') \approx 1/4 \pi \epsilon_0 |\vect{r} -  \vect{r}'|$ when contacts are sufficiently far away and 
capping layers are sufficiently thin. Now we write down the induced charge density in the sample in terms of the density response function, 
\begin{equation} \label{eq:formal}
    \delta n (\vect{r},\omega) = e \int \dd \vect{r}' \rchi(\vect{r},\vect{r}';\omega) V_{\mathrm{ext}}(\vect{r}',\omega)
\end{equation}
with $\vect{r}$ and $\vect{r}'$ both inside the sample. We will drop the index $\omega$ below to simplify the notation. This induced charge density in turn generates a potential $V_t = - e \int \dd \vect{r} G(\vect{r}_t, \vect{r}) \delta n(\vect{r})$ at the apex of tip. The total potential at the peak of tip has an additional term due to the tip head capacitance $C_t$, $V_{\mathrm{tot}} = Q_t / C_t + V_t$, then we can find the tip--sample admittance by expanding $Y_{\mathrm{tot}} = \partial_t Q_t(t) / V_{\mathrm{tot}}$ to the leading order in $C_t V_t/Q_t$,
\begin{align} 
    Y_{\ts} \approx - \ii \omega e^2 C_t^2 \int \dd \vect{r} \dd \vect{r}' G(\vect{r}_t, \vect{r}) \rchi(\vect{r},\vect{r}') G(\vect{r}', \vect{r}_t) \label{eq:MIM}
\end{align}
We note that all constant terms and prefactors will be absorbed in the constants $a$ and $b$ in Eqn.~(\ref{eq:ab}) during impedance matching and therefore neglected from now on. Eqn.~(\ref{eq:MIM}) is one of the main results of this work, which characterizes the MIM response of arbitrary quantum systems {with $\chi(\vect{r},\vect{r}')$ describing the dynamical charge density-density correlation between $\vect{r}$ and $\vect{r'}$ in the sample}. The derivation of this equation uses only two assumptions: the tip can be approximated by a single point, and the microwave frequency electromagnetic field generated by MIM can be treated within linear response theory. The first assumption will be verified in a special case through numerical simulations in Sec.~\ref{sec:simulation}, while the second assumption is supported by the high sensitivity of MIM provided by impedance matching.

Before we proceed to apply this general framework to Chern insulators, we will make a few comments on the result presented in Eqn.~(\ref{eq:MIM}). First, this general result reduces to the lumped element model in the case of a simple homogeneous metal sheet with dielectric function $\epsilon(\bq)$,
\begin{equation}
\begin{gathered}
    \rchi(\bq) = \frac{\epsilon_0^2}{e^2} \left ( \frac{1}{\epsilon_0} - \frac{1}{\epsilon(\bq)} \right ) |\bq|^2\\
    Y_{\ts} \sim - \ii \omega \frac{1}{32 \pi} \frac{t^2}{d^2} \left ( \frac{1}{\epsilon_0} - \frac{1}{\epsilon_{\mathrm{eff}}} \right )
\end{gathered}
\end{equation}
where $t$ is the sample thickness, $d$ is the tip--sample distance, and $\epsilon_{\mathrm{eff}}$ is the effective dielectric constant. $\epsilon_{\mathrm{eff}}$ approaches to the usual DC dielectric constant $\epsilon(\bq = 0)$ in the limit when the tip is sufficiently far from the sample. The lumped element model parameters are given by sample resistance $R_s \sim 1/\omega \Im \epsilon_{\mathrm{eff}}$, sample capacitance $C_s \sim \Re \epsilon_{\mathrm{eff}}$. The tip--sample capacitance $C_0$ that only shows up in constant terms and prefactors (see Ref.~\onlinecite{SM} Sec.~\ref{sec:reduce} for definitions and a full derivation). 

Secondly, in the limit when the tip is sufficiently far from the sample, we can further approximate $\rchi(\bq)$ by $\rchi(\bq = 0)$; then the MIM signal can be related to the electronic compressibility $\dd n/\dd \mu$ as follows:
\begin{equation}
    Y_{\ts} \sim - \ii \rchi(\bq = 0) = - \ii \frac{\delta n}{\delta \mu}
\end{equation}
where $n$ is the electron density and $\mu$ is the chemical potential. This provides a direct correspondence between the imaginary part of the MIM response and the chemical potential $\mu(n)$ (which for example can be measured by scanning single-electron transistors and related field penetration techniques \cite{Ben1, Ben2}). 

Finally, due to the factor of $i$ in Eqn.~(\ref{eq:MIM}), the imaginary part of the MIM response, $\Im Y_{\ts}$, comes from the real part of the density response function, $\Re \rchi$, and vice versa. Thus, $\Im Y_{\ts}$ and $\Re Y_{\ts}$ measure the reflectivity and absorption of the sample, respectively.

\subsection{Special case: Application to QAH insulators}

To illustrate a concrete example, we now apply this general framework to calculate the MIM response of topological edge modes in a QAH insulator. The recipe is to first find all low energy excitations, then compute these excitations' contribution to the density response function $\rchi$, and finally apply Eqn.~(\ref{eq:MIM}) to find the MIM response.
In two-dimensional QAH insulators, the bulk contributes little to the dielectric response because of the large energy gap, so we only need to focus on the edge contribution. 
The edge states in QAH insulators are characterized by the chiral Luttinger liquid that host plasmonic excitations \cite{chiLL,Chetan,MacDonaldEMP}.
These gapless edge plasmons modes are called edge magnetoplasmons (EMP) in the context of quantum Hall insulators and have been studied extensively for several decades \cite{MacDonaldEMP,EMPExpt1,EMPExpt2,VolkovEMP,Justin,Liang,Chetan}. More recently, they have also been studied in the context of QAH insulators \cite{DGG}, quantum spin Hall insulators \cite{ErwannEMP} and Chern insulators \cite{Justin}. To fully understand EMP modes beyond the chiral Luttinger liquid, prior theory work has also considered semiclassical corrections \cite{VolkovEMP, Liang,MacDonaldEMP, Justin}. 

In addition to EMP modes, edge acoustic modes have also been observed in the low energy excitation spectrum in QAH insulators \cite{AG,CVSpectral}. However, these modes are overdamped at microwave frequencies in the parameter regime of QAH insulators. Therefore, we can safely conclude that the MIM response in QAH insulators is dominated by edge plasmon modes. This is the underlying reason why the MIM response of a QAH insulator behaves so differently from the response of materials in which electron-hole pair excitation dominates. It might be a bit surprising since plasmon modes are usually not expected to show up in MIM data due to their finite energy gap. However, in QAH insulators these EMP modes are gapless due to the 1D nature of the edge states. In the following, we will only focus on the EMP contribution to $Y_{\ts}$.

If we place the tip a linear distance $r$ from the edge, the tip-sample admittance reduces to
\begin{equation} \label{eq:ysimplify}
    Y_\text{ts} \sim - \frac{\ii \omega e^2}P \sum_k G_r(k) \rchi(k) G_r(-k)
\end{equation}
where $k$ is the 1D momentum along the edge, $P$ is the sample perimeter, and $G_r(k) = K_0(|k| r)/4 \pi \epsilon_0$ is the 1D Fourier transform of the Coulomb interaction $1/4 \pi \epsilon_0 |\vect{r} -  \vect{r}'|$ with $K_0$ being the modified Bessel function. In Ref.~\onlinecite{SM} Sec.~SII, we compute $\rchi(k)$ for chiral edge modes at the RPA level,
\begin{equation} \label{eq:chichiralmain}
    \rchi(k) = \frac{1}{2\pi}  \frac{k}{\hbar \omega - \hbar v k - \frac{e^2}{(2 \pi)^2 \epsilon_{0}} k \log \left( \frac{1}{k l} \right)}
\end{equation}
where $v$ is the edge velocity and $l$ is the localization length of the edge mode, which is inversely proportional to the bulk gap and is assumed to be small compared to $r$ in Eqn.~(\ref{eq:ysimplify}). In pure QAH samples where the Dirac velocity $v_F$ is the only energy scale, we expect $v = v_F$. The poles of $\rchi(k)$ define the EMP frequencies,
\begin{equation} \label{eq:EMP}
    \hbar \omega_{\mathrm{EMP}}(k) =  \hbar v k + \frac{e^2}{(2 \pi)^2 \epsilon_{0}} k \log \left( \frac{1}{k l} \right)
\end{equation}
This result agrees with Ref.~\cite{MacDonaldEMP}, where the first term is referred to as the quantum part, and the second term is referred to as the classical part. 
The imaginary part of $\rchi(k)$ can be obtained by taking $\omega \mapsto \omega+ \ii \epsilon$, then $\Im \rchi(k,\omega+\ii\epsilon) = - k \delta(\hbar \omega - \hbar \omega_{\mathrm{EMP}}) / 2$ with $\delta$ being the delta function. In the experiment, $\epsilon$ could be a small but finite number due to dissipation effects, then $\Im \rchi(k)$ would become a Lorentzian function centered at $\hbar \omega_{\mathrm{EMP}}$ (see Ref.~\onlinecite{SM} Sec.~\ref{sec:dissipation}). 
Therefore, we expect the lineshape of the EMP resonance peaks to be broadened in the presence of dissipation along the sample boundaries. These features should allow microwave imaging to shed light on the the microscopic nature of the dissipation in the chiral edge modes, as revealed by the MIM response in the frequency domain.

Since we are operating at microwave frequencies, we have to worry about the quantization of momentum $k = 2 \pi n/P$ with $n$ being an integer, given the finite sample diameter $P$. If the MIM frequency $\omega \sim \omega_{\mathrm{EMP}}(2 \pi/P)$ is comparable to the lowest few EMP resonances, the dominant contribution to the MIM response comes from a particular momentum $k_{\omega}$ whose EMP frequency $\omega_{\mathrm{EMP}}(k_{\omega})$ is closest to $\omega$, then 
\begin{equation} \label{eq:MIMQAH}
    Y_\text{ts} \sim - \frac{\ii \omega e^2}{32 \pi^3 \epsilon_0^2 P} \frac{ k_{\omega}}{\hbar \omega - \hbar \omega_{\mathrm{EMP}}(k_{\omega})} K_0^2(k_{\omega} r) \underset{r \to 0}{\sim} - \log^2 k_{\omega} r
\end{equation}
In experiments, we expect the case $\omega \sim \hbar \omega_{\mathrm{EMP}}(2 \pi/P)$ to hold given the size of the system \cite{DGG}. 

The presence of edge magnetoplasmons should be manifested experimentally in a strong enhancement of the MIM response at the boundaries of the QAH insulator, in agreement with prior experimental results \cite{Monica,JianQAH}. Fig.~\ref{fig:MIMTheory} (b) shows the spatial profile of the MIM response ($\Im \MIM \sim \Im Y_\text{ts}$) of a QAH edge mode, revealing a sharp peak at the crystal boundaries, following Eqn.~(\ref{eq:MIMQAH}). 
The width of this edge peak is given by a characteristic length scale  $1/k_{\omega}$, the EMP wavelength, 
which can be as large as a few microns for realistic sample dimensions. It is generally expected that the edge width measured with MIM will be larger than the actual edge localization length $l$ due to long range Coulomb coupling between the tip and the sample. 
This observation helps explain the vastly different edge state widths reported in transport \cite{WangLocalization,ChangLocalization} and STM \cite{STM1,STM2,STM3,STM4,STM5} studies versus MIM experiments \cite{Monica,JianQAH}. 


Investigating the frequency dependence of the MIM edge peak should shed light on the topological nature of the Chern insulator state. 
Fig.~\ref{fig:MIMTheory} (c) illustrates the evolution of the edge peak amplitide as a function of the MIM excitation frequency  $\omega$. Here we plot $\Im \MIM$ as a function of $\omega$ for a fixed tip location over the sample edge, with $d = 0.005 P$, $l = 0.2 P$, and $e^2/((2\pi)^2 \epsilon_0) = 0.2 \hbar v$. 
The EMP resonances appear at a series of discrete microwave frequencies, which are found to depend quantitatively on both the topological invariant and the sample perimeter (see Ref.~\onlinecite{SM} Sec.~\ref{sec:reduce}) \cite{DGG,MacDonaldEMP,EMPExpt1,EMPExpt2}. Because trivial edge states localized at crystal boundaries should be featureless as a function of frequency, this unique fingerprint of the EMP resonances in the frequency domain provides a route to unambiguously differentiate between topological and trivial edge modes. 
The quantitative relationship between the resonance frequency and the sample circumference also illustrates non-local, topological nature of the chiral edge modes that circulate around the entire sample.

In practice, because continuously sweeping the MIM frequency can be experimentally challenging, one could first identify the EMP resonances using traditional microwave transmission measurements \cite{DGG} and then perform MIM imagining at a few frequencies close to and away from those EMP resonances.


\begin{figure*}[htbp]
    \centering
    \includegraphics[width=0.9\textwidth]{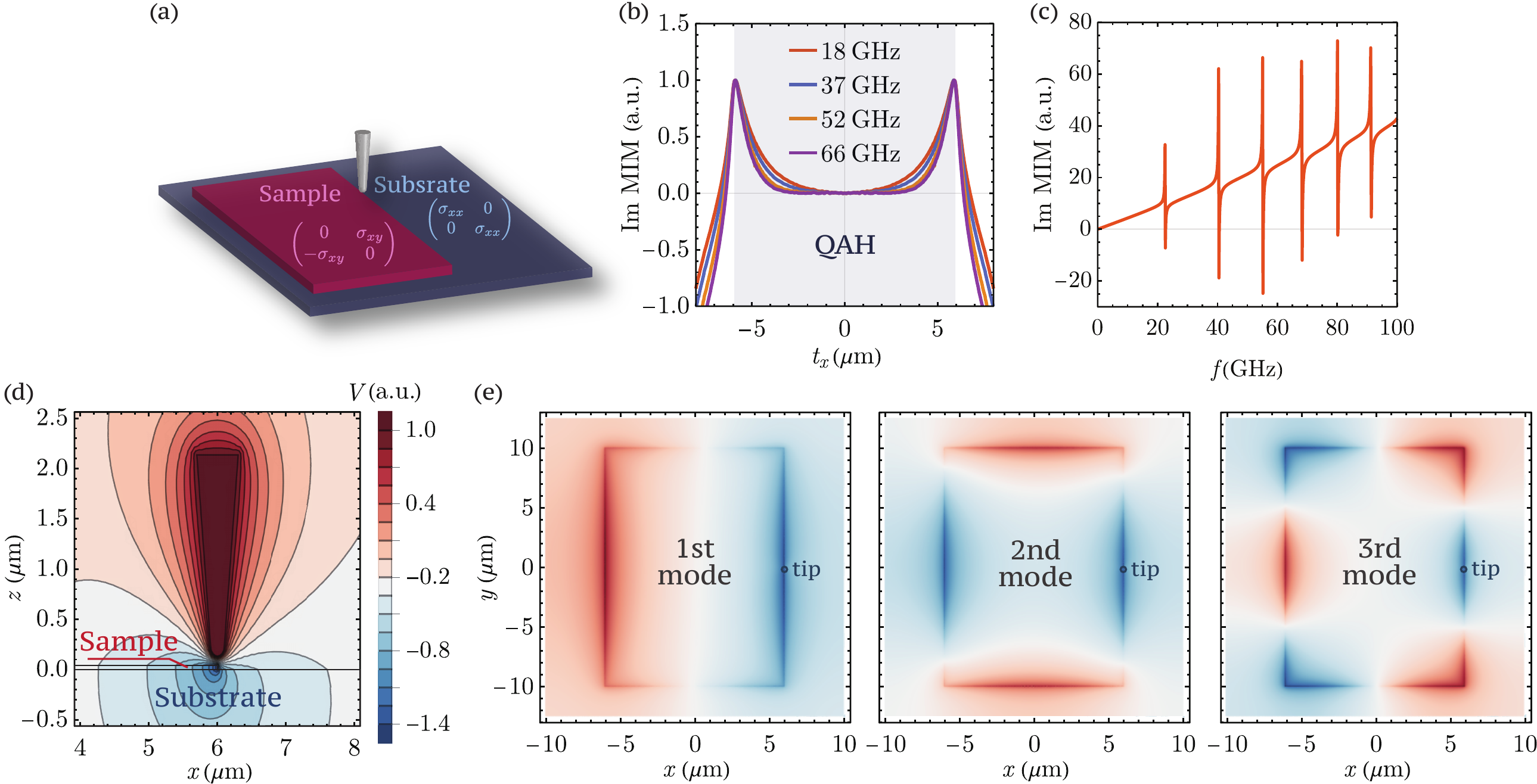}
    \caption{\textbf{Numerical simulation of topological edge magnetoplasmons in a QAH insulator.} \textbf{(a)} Schematic illustration of the tip-sample  setup used in the numerical simulation. The sample is characterized by a non-zero Hall conductance $\sigma_{xy} = e^2/h$ and a vanishing longitudinal conductance $\sigma_{xx}$. \textbf{(b)} Imaginary part of the MIM response, measured as a function of position across the sample, at various frequencies. A strong enhancement of the MIM response is observed at the sample boundaries due to the presence of edge magnetoplasmon (EMP) modes. (\textit{Note:} Here we work with frequencies much higher than usually used in MIM experiments to compensate the fact that the EMP frequencies are scaled up when we scale down the sample dimension to ensure numerical convergence.) \textbf{(c)} Imaginary part of the MIM response, plotted as a function of frequency when the tip is positioned over the sample edge. The difference between this simulation result and the analytical result in Fig.~\ref{fig:MIMTheory} mainly comes from the dielectric environment. \textbf{(d-e)} Electric potential distribution in (d) the vertical plane across the tip and (e) the horizontal plane across the sample at the first few EMP resonances. The location of the tip is marked, which pins the phase of the edge charge distribution (see main text).}
    \label{fig:MIMSimu}
\end{figure*}

\section{Semiclassical simulations}
\label{sec:simulation}

To verify our analytical results, which were derived using an approximation that treats the MIM tip as a point, we also perform a numerical simulation of the MIM signal that incorporates a realistic tip-sample geometry and dielectric environment that had been neglected in the analytical treatment. The simulation still focuses on the topological edge state contribution to the MIM signal and utilizes a classical formulation of the EMP modes, which accurately capture the density response of these modes even though it may not accurately reproduce the EMP frequencies \cite{VolkovEMP,Justin}. This formulation requires solving Maxwell's equations with a non-zero Hall conductance inside the sample, which therefore only captures the classical part of the EMP frequencies in Eqn.~(\ref{eq:EMP}).

To compute the admittance $Y_{\mathrm{ts}}$, we use finite element analysis to numerically solve Maxwell's equations for the entire experimental setup, including the MIM tip, the sample, and the substrate, as illustrated schematically in Fig.~\ref{fig:MIMSimu} (a) (see Ref.~\onlinecite{SM} Sec.~\ref{sec:simapp} for details). 
The sample is placed on top of the substrate as in the experiment \cite{Monica}, which results in a step across the sample boundary (Fig.~\ref{fig:MIMSimu} (a)). The dielectric environment is set to be identical to the experiment setup, and the conductivity tensor is set to be $\sigma_{xy} = e^2/h$ inside the sample and zero outside. To ensure numerical convergence, we set the sample size to $\SI{12}{\um} \times \SI{20}{\um}$, which is much smaller than the one used in the experiment in Ref.~\onlinecite{Monica} and Sec.~\ref{sec:expt}. A side effect of scaling down the sample dimension is that we need to look at much higher frequencies in order to compare with the experiment since the first few EMP frequencies $\omega_n \sim \hbar \omega_{\mathrm{EMP}}(2 \pi n/P)$ are scaled up the same time (see Eqn.~(\ref{eq:EMP}) and Ref.~\onlinecite{SM} Sec.~\ref{sec:frequency} for details). More quantitatively, the first few EMP frequencies are about $13$ times larger than one would expect in the sample to be discussed in Sec.~\ref{sec:expt} given their difference in perimeter. 
The topological nature of the problem requires a careful choice of the solver to ensure convergence \cite{SM}.

Fig.~\ref{fig:MIMSimu} (b) displays a real-space plot of $\Im Y_{\mathrm{ts}}$ as the tip is scanned across the sample, which reproduces clear peaks at the sample boundaries, $x = \pm \SI{6}{\um}$. The spatial profile of these edge peaks, including the decay length inside the sample, agree quantitatively with the analytical predictions in Fig.~\ref{fig:MIMTheory} (b). The most obvious difference lies in the asymmetry of $\Im Y_{\mathrm{ts}}$ inside and outside the sample in the simulation, which comes from the step and a change in dielectric environment across the sample boundary.

In Fig.~\ref{fig:MIMSimu} (b), we note that the spatially-resolved MIM signal is plotted at a series of generic frequencies, which are typically away from exact resonance frequencies. 
Upon analyzing the frequency dependence of $\Im Y_{\mathrm{ts}}$, we find the edge peak to be narrower at higher frequencies. This phenomenon has a simple explanation within our theoretical framework. From Eqn.~(\ref{eq:EMP}), we know that a higher frequency $\omega$ corresponds to a larger EMP momentum $k_{\omega}$. Meanwhile, $1/k_{\omega}$ sets the decay length scale of the MIM signal when the tip moves away from the edge (see Eqn.~(\ref{eq:MIMQAH})). Combining these two observations, the peak is expected to be narrower at higher frequencies at which the MIM response comes from a higher frequency EMP mode associated with a shorter decay length. This prediction will later be compared with experiments in the following section.

When the tip is positioned over the sample edge, the imaginary part of $Y_{\mathrm{ts}}$ picks up a series of resonance peaks corresponding to the first few EMP modes (Fig.~\ref{fig:MIMSimu} (c)), in agreement with analytical predictions. 
However, compared to Fig.~\ref{fig:MIMTheory} (b), there is an additional contribution from the dielectric environment that scales linearly with the MIM frequency. Another noticeable feature is the absence of a zero frequency peak due to the factor of $k_{\omega}$ in Eqn.~(\ref{eq:MIMQAH}). In Ref.~\onlinecite{SM} Sec.~SIV, we extract the dispersion of the EMP modes by identifying resonance peaks in $\Im Y_{\mathrm{ts}}$, which agrees quantitatively well with the classical part of Eqn.~(\ref{eq:EMP}).

Fig.~\ref{fig:MIMSimu} (d-e) illustrates the electric potential distribution 
in the plane of the sample at the first few EMP frequencies, which provides a visualization of the real-space charge density oscillations at the fundamental EMP mode and higher harmonics. As shown in Fig.~\ref{fig:MIMSimu} (e), when the MIM frequency $\omega$ coincides with these EMP resonance frequencies, positive and negative charges start to concentrate at the sample edge with an in-plane distribution $\delta n \sim \delta (r_{\bot}) e^{i(k r_{\parallel} + \phi)}$, where $r_{\parallel}$ is the distance along the edge,  $r_{\bot}$ is the distance from the edge. We note that the phase $\phi$ is pinned by the location of the tip since the energy is minimized when the charge distribution is most negative beneath the tip. The characteristic potential distribution of standing wave patterns clearly identifies their plasmonic nature, while also confirming that the experimental setup is able to excite EMP modes. In Fig.~\ref{fig:MIMSimu} (d), we see that the potential starts to decay immediately away from the edge, suggesting that charges are spatially confined to the boundaries of the sample.

Finally, we comment on the how expected MIM signatures of the EMP modes in a QAH insulator can be distinguished from the signatures of trivial edge modes arising from impurities at the boundaries of a conventional insulator. As shown in Ref.~\onlinecite{SM} Sec.~\ref{sec:non-QAH}, the peak conductance required to reproduce the shape of the MIM curve is as high as $\SI{1e7}{S/m}$, which would require a large concentration of metallic impurities that are extremely unlikely given the current fabrication process. Additionally, the peak in the MIM response at the boundaries of the sample is found to vanish at magnetic fields corresponding to the $\sigma_{xy} = 0$ phase, which suggests that the enhanced edge conduction not trivial in nature\cite{Monica}. Meanwhile, in the trivial case, the profile of the MIM edge peak would remain the same across a large range of frequencies, in contrast to the expected MIM response from EMP modes based on the theoretical picture presented above.


\begin{figure*}[htbp]
    \centering
     \includegraphics[width = 0.98\textwidth]{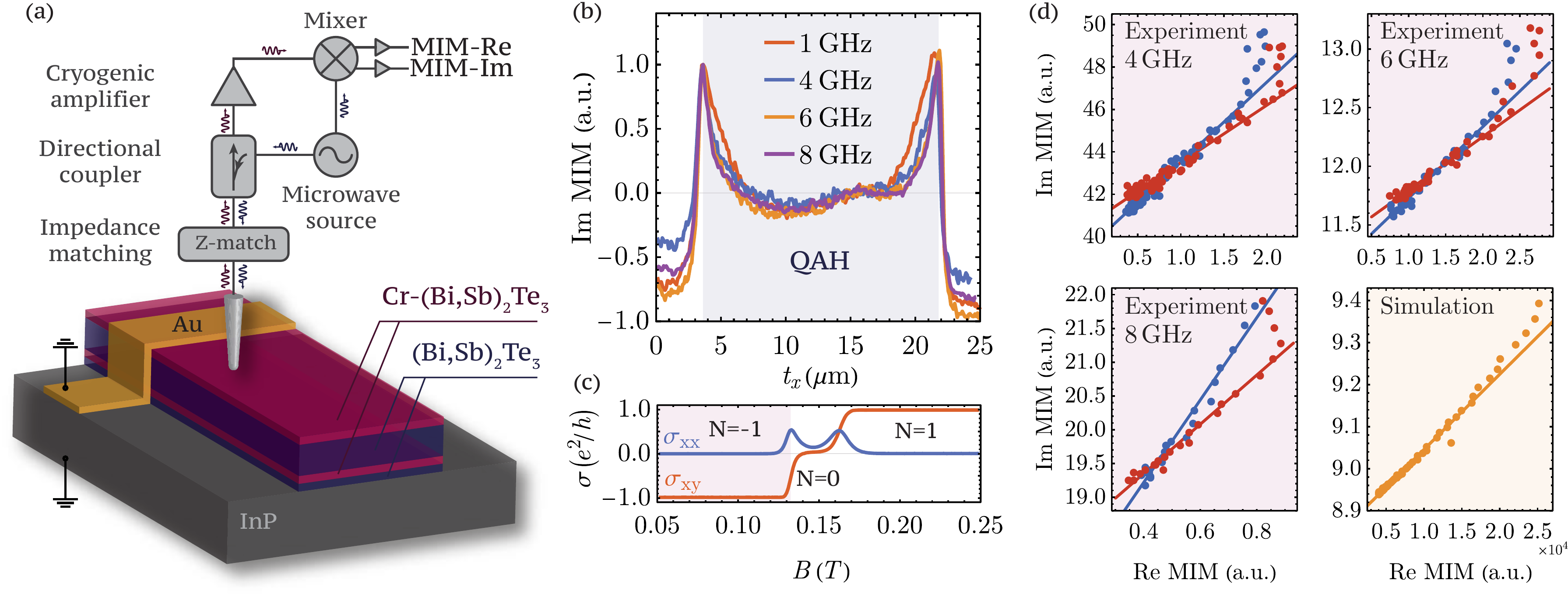}
    \caption{\textbf{Spatially-resolved measurement of chiral edge states in a quantum anomalous Hall insulator.} \textbf{(a)} Schematic illustration of the MIM experimental measurement setup. The tip is scanned across the sample, parallel to the shorter edge. \textbf{(b)} Experimental data reveals a strong enhancement of the imaginary part MIM response at the edge of the sample in the quantum anomalous Hall (QAH) regime (the sample lies within the shaded region). The measured edge peak profile becomes narrower as the MIM frequency increases. \textbf{(c)} Transport characterization of the sample, showing a well-quantized Hall conductance and vanishing longitudinal conductance in the QAH phase. \textbf{(d)} The $\Re \MIM$ vs. $ \Im \MIM$ scatter plot of the experimental data at \SI{4}{\GHz}, \SI{6}{\GHz}, and \SI{8}{\GHz} and the simulation results around the first EMP frequency. The red (blue) curves correspond to data close to the left (right) sample edge.}
    \label{fig:MIMExpt}
\end{figure*}

\section{Experimental results} \label{sec:expt}

To verify our theoretical framework, we performed real-space microwave imaging of one-dimensional QAH edge states in a high-quality magnetic topological insulator thin film at various frequencies. The experimental setup is shown in Fig.~\ref{fig:MIMExpt} (a). The sample is a \ce{Cr}-doped \ce{(Bi,Sb)2Te3} Hall bar with dimensions of \SI{400}{um} by \SI{20}{um}, and we scan the MIM tip across the device parallel to the short edge. Details on device preparation and the MIM measurement setup can be found in Ref.~\onlinecite{SM} Sec.~\ref{sec:divapp}. Transport measurements are used for a baseline characterization of the quality of the QAH insulator state. As shown in Fig.~\ref{fig:MIMExpt} (c), the Hall conductance $\sigma_{xy}$ is fully quantized and the longitudinal conductance $\sigma_{xx}$ drops to zero at $B = \SI{50}{mT}$ below the coercive field, suggesting that current is primarily transmitted by chiral edge modes that are topologically protected from backscattering in the QAH phases (labeled by Chern number $N=\pm 1$).

The presence of the QAH edge modes is manifested experimentally in a sharp enhancement of the MIM response at the boundaries of the sample, as shown in the spatially-resolved microwave imaging data presented in Fig.~\ref{fig:MIMExpt} (b). 
(We remind readers that the experiment and the numerical simulations were performed at very different frequencies to compensate the difference in sample dimensions, but the resulting MIM data can still be compared at a qualitative level.)

These experimental results have a few surprising features that differ from the expected behavior of a conventional insulator with an enhanced edge conductivity with trivial origins.
First of all, the observed MIM signal is much stronger than that expected from edge defects or local doping (see Ref.~\onlinecite{SM} Sec.~\ref{sec:non-QAH} for a comparison). If the MIM signal comes from the EMP modes, however, it is expected to diverge at EMP frequencies and can therefore be large in general. Another feature is that the real part of the measured MIM response is much smaller than the imaginary part, as shown in Fig.~\ref{fig:MIMExpt} (d). This can be explained by Eqn.~(\ref{eq:chichiralmain}) since $\Re \rchi(\bk)$ is tiny except when the MIM frequency precisely hits one of the EMP frequencies.
In addition, the spatial profile of the MIM response near the sample edge becomes narrower at a higher frequencies, which agrees qualitatively with the numerical simulations (see Fig.~\ref{fig:MIMExpt} (b) and Fig.~\ref{fig:MIMSimu} (b)). 

Finally, we also investigate the relationship between the real and imaginary parts of the MIM response, measured as a function of position, at various frequencies. As shown in Fig.~\ref{fig:MIMExpt} (d), we note that the real and imaginary parts of the observed MIM response have a linear relation at frequencies higher than \SI{4}{GHz}. This linear relationship provides strong evidence in favor of EMP modes being responsible for the enhanced MIM signal at the sample edge, and stands in sharp contrast to the expected semi-circle relation predicted by the lumped-element model. The results suggest that the MIM signal mainly comes from a 1D edge, then the $r$ dependence of $\Re \MIM$ and $\Im \MIM$ has to be the same. At frequencies lower than the first EMP resonance, the interpretation is complicated by the dielectric background. We refer interested readers to Ref.~\onlinecite{SM} Sec.~\ref{sec:scatter} for more details.

\section{Discussion and Outlook}

This paper provides the first quantitative interpretation of the MIM response of quantum materials within linear response theory. 
In the limit when the tip is sufficiently far from the sample, we  show that the imaginary part of the MIM response can be quantitatively related to the electronic compressibility. We would like to take this opportunity to compare MIM and the scanning SET technique, which directly measures chemical potential and therefore the electronic compressibility\cite{Ben1, Ben2}. MIM has the advantage of being less constrained by the electrostatic gating setup (and the associated fringing fields near the sample boundaries) and has a higher spatial resolution due to a simpler tip geometry, while scanning SET has the benefit of providing a more quantitative measurement of gap sizes and electronic compressibility without the need for impedance matching.

To illustrate a concrete application of the general model above, we compute the MIM response of a QAH insulator and reveal that the experimentally observed enhancement of the MIM signal at the sample boundaries comes from topological edge magnetoplasmon modes. 
This observation allows one to experimentally distinguish topologically nontrivial from trivial edge modes by investigating the quantitative relationship between the real and imaginary parts of the complex MIM response at multiple frequencies. 
Furthermore, this theoretical picture also allows us to predict the width $1/k_{\omega}$ of the experimentally observed peak of in MIM response at the boundaries of Chern insulators, which explains the apparent inconsistencies between the edge state decay length scales measured using MIM versus STM or transport\cite{ShenHall,PhysRevLett.117.186601,PhysRevB.86.195417,Kim2021,PhysRevB.107.115426}.

To confirm and expand on the analytical results, we performed numerical simulations that took into account the effects of the tip-sample geometry and the dielectric background: the former turned out to be a small correction and the later only added linearly to the MIM signal. 
We also performed MIM measurements of the Chern insulator states in a  \ce{Cr}-doped \ce{(Bi,Sb)2Te3} magnetic topological insulator at multiple \SI{}{\GHz} frequencies to verify our theoretical understanding. We observed a clear peak in the MIM response at the edge of the sample, whose spatial profile, frequency dependence, and ratio of the real and imaginary parts of the MIM signal were consistent with our framework. We would like to point out that future MIM experiments with continuous frequency tunability would be desirable to fully verify our EMP interpretation of the MIM edge response in QAH insulators.

\begin{figure}[htbp]
    \centering
    \includegraphics[width=0.45\textwidth]{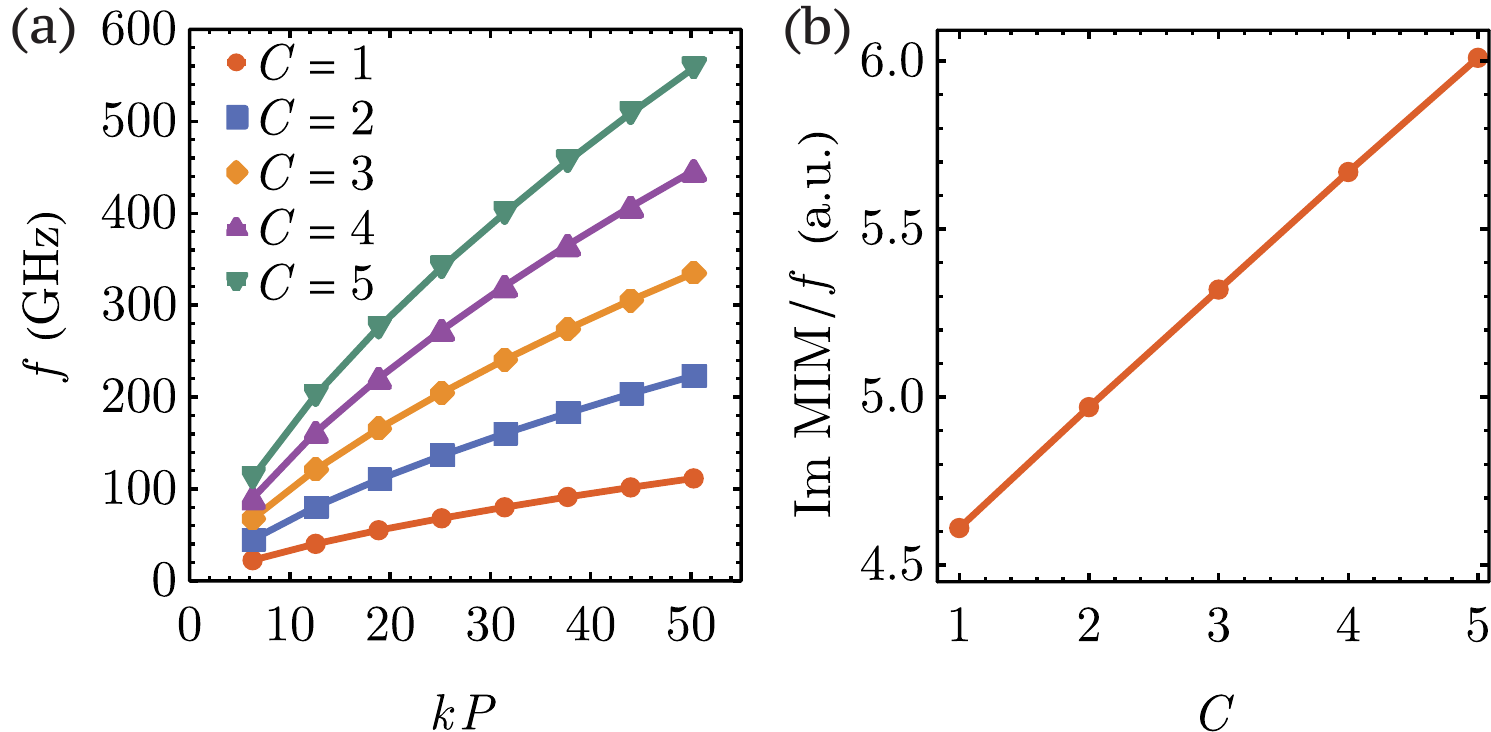}
    \caption{\textbf{Numerically simulated MIM response of a higher Chern number insulator.} Here we present numerical simulations of the MIM response of a Chern insulator with the tip positioned directly over the edge state. The topological invariant, or Chern number, is $C$. \textbf{(a)} Plot of the EMP resonance frequencies for a variety of Chern insulator states with different integers $C$. The curves can be fitted by the same parameters of $\epsilon_r$ and $l$ as in Fig.~\ref{fig:MIMSimu} (c) inset. \textbf{(b)} Imaginary part of the MIM signal at EMP frequency $f$, plotted as a function of Chern number. Here we consider a finite dissipation $\sigma_{xx} \neq 0$ and factor out the trivial factor $f$ in Eqn.~\eqref{eq:formal} for clarity. The MIM signal has a finite magnitude at $C = 0$ due to baseline contributions from the dielectric environment. }
    \label{fig:Chern}
\end{figure}

Finally, we would like to comment on our new formulation's predictions for measuring the topological invariant of a Chern insulator using the technique of MIM. Expanding upon our previous calculations for QAH insulators with Chern number $\nu = 1$, the resonance frequency and MIM response can also be computed for systems with higher Chern numbers by including multiple branches in the density response calculation. At classical level, we expect both the EMP frequencies and the MIM signal magnitude right on the resonance to be linearly proportional to the Chern number as shown in Fig.~\ref{fig:Chern},
$\hbar \omega_{\mathrm{EMP}} \propto C, \quad Y_{\mathrm{ts}} \propto C$. 
We refer readers to Ref.~\onlinecite{SM} Eqn.~\eqref{eq:C} for the full analytical formula with quantum corrections. 
We also expect the lineshape of the MIM signal to depend on the dissipation along the chiral edge states (manifested in a finite $\sigma_{xx}$), as discussed in Section II. These features should allow the technique of MIM to shed light on the Chern number of a topological state, as well as the microscopic nature of the dissipation at the sample boundaries.

\begin{acknowledgments}
We thank Allan H. MacDonald, Tarun Grover, Yongtao Cui, Rahul Roy, Massoud R. Masir, Yahya Alavirad and Adrian B. Culver for inspiring discussions. We thank Alex Cauchon for helping develop some of the numerical simulations. 
M.T.A acknowledges funding support from the UC Office of the President, specifically the UC Laboratory Fees Research Program (award LFR-20-653926) and the AFOSR Young Investigator Program (award FA9550-20-1-0035).
T.W. is supported by the U.S. Department of Energy, Office of Science, Office of Basic Energy Sciences, Materials Sciences and Engineering Division under Contract No. DE-AC02-05-CH11231 (Theory of Materials program KC2301). This work was partly supported by Japan Science and Technology Agency Core Research for Evolutional Science and Technology (JPMJCR16F1).  Y.Z.Y. is supported by the NSF Grant DMR-2238360. Microwave impedance microscopy measurements at Stanford University were supported by the Gordon and Betty Moore Foundation’s Emergent Phenomena in Quantum Systems initiative through Grant GBMF4546 (Z.-X.S.) and by the National Science Foundation through Grant DMR-1305731. 

\end{acknowledgments}

\bibliography{nc_v1.bib}

\onecolumngrid

\vspace{0.3cm}

\newpage
\begin{center}
\Large{\bf Supplementary Information}
\end{center}

\supplementarysection

\section{Relation between the general framework and the lumped-element model} \label{sec:reduce}

In this section, we want to show that the general framework reduces to the lumped-element model in the case of a simple homogeneous metal. To do so, we first review the connection between dielectric function, density response function, and the conductivity. In this calculation, we will work with a three dimensional isotropic sample, which applies when the Thomas-Fermi screening length $\lambda_{\mathrm{TF}}$ is shorter than the sample thickness $t$. The density response function $\rchi(\bq)$ is defined as
\begin{equation} \label{eq:chidef}
    \delta n (\bq) = e \rchi(\bq) V_{\mathrm{ext}}(\bq) = \frac{e}{\epsilon_0 |\bq|^2} \rchi(\bq) Q_{\mathrm{ext}}(\bq)
\end{equation}
where $Q_{\mathrm{ext}}$ is the external charge, $\delta n$ is the induced number density, and $G(\bq) = 1/\epsilon_0 |\bq|^2$ is the Coulomb interaction. Using Gauss's law, we find
\begin{equation} \label{eq:gauss}
    i \bq \cdot \vect{D}(\bq) = Q_{\mathrm{ext}}(\bq), \quad i \bq \cdot \vect{E}(\bq) = \frac{1}{\epsilon_0} (Q_{\mathrm{ext}} - e \delta n (\bq))
\end{equation}
The dielectric function $\epsilon(\bq)$ is defined as
\begin{equation} \label{eq:chi}
    \frac{1}{\epsilon(\bq)} = \frac{\vect{E}(\bq)}{\vect{D}(\bq)} = \frac{Q_{\mathrm{ext}} - e \delta n (\bq)}{\epsilon_0 Q_{\mathrm{ext}}(\bq)} = \frac{1}{\epsilon_0} - \frac{e^2}{\epsilon_0^2 |\bq|^2} \rchi(\bq)
\end{equation}
then the response function can be related to $\epsilon(\bq)$ by 
\begin{equation}
    \rchi(\bq) = \frac{\epsilon_0^2}{e^2} \left ( \frac{1}{\epsilon_0} - \frac{1}{\epsilon(\bq)} \right ) |\bq|^2
\end{equation}
To related the response function to conductivity, we first write down the continuity equation,
\begin{equation} \label{eq:continuity}
    - i e \omega \delta n(\bq) = i \bq \cdot \vect{j}(\bq) = i \sigma(\bq) \bq \cdot \vect{E}(\bq)
\end{equation}
where $\vect{j}$ is the electrical current. Combining Eqn.~(\ref{eq:gauss}) and Eqn.~(\ref{eq:continuity}) and plugging in Eqn.~(\ref{eq:chidef}), we find
\begin{gather}
    - i e \omega \delta n(\bq) = \frac{\sigma(\bq)}{\epsilon_0 } (Q_{\mathrm{ext}}(\bq) - e \delta n (\bq))\\
    i \omega = \frac{\sigma(\bq)}{\epsilon_0} \frac{\epsilon_0}{\epsilon_0-\epsilon(\bq)}\\
    \epsilon(\bq) = \epsilon_0 + i \frac{\sigma(\bq)}{\omega} \label{eq:diec}
\end{gather}

Now we can evaluate the MIM response using Eqn.~(\ref{eq:MIM}). In the limit $d$ is much larger than the sample thickness $t$, we can approximate $G(\vect{r}_t,\vect{r}) \approx G(\vect{r}_{\parallel}) = 1/4\pi \epsilon_0 (|\vect{r}_{\parallel}|^2+ d^2)^{1/2}$, and $\rchi(\vect{r},\vect{r}') \approx \rchi(\vect{r}_{\parallel}-\vect{r}'_{\parallel})$,
\begin{align} \label{eq:MIM3D}
    Y_{\rchi}
    &\sim - i \omega e^2 \int \dd \vect{r} \dd \vect{r}' G(\vect{r}_{\parallel}) \rchi ( \vect{r}_{\parallel}-\vect{r}'_{\parallel} ) G(\vect{r}'_{\parallel})\\
    &\sim - i \omega e^2 t^2 \int \frac{\dd \vect{q}}{(2\pi)^2} G(\vect{q}) \rchi(\vect{q}) G(-\vect{q})\\
    &\sim - i \omega \frac{1}{32 \pi} \frac{t^2}{d^2} \left ( \frac{1}{\epsilon_0} - \frac{1}{\epsilon_{\mathrm{eff}}} \right )
\end{align}
where $\bq$ is the two dimensional momentum, the Fourier transform of $G(\vect{r}_{\parallel})$ takes the form $G(\vect{q}) = e^{-|\vect{q}| d}/2 \epsilon_0 |\vect{q}|$, and
\begin{equation}
    \frac{1}{\epsilon_{\mathrm{eff}}} \equiv 8 \pi d^2 \int \frac{\dd \vect{q}}{(2\pi)^2} \frac{e^{-2|\vect{q}| d}}{\epsilon(\bq)}
\end{equation}
The approximation $\rchi(\vect{r},\vect{r}') \approx \rchi(\vect{r}_{\parallel}-\vect{r}'_{\parallel})$ is justified since the error mainly comes from $|\bq| \gg 2\pi/t$, which is exponentially suppressed by $G(\vect{q})$ in the limit $d \gg t$. 
If $\epsilon(\bq)$ does not vary significantly in the regime $|\bq| \ll 2\pi/d$, we notice that $\epsilon_{\mathrm{eff}} \approx \epsilon(\bq = 0)$ which is the DC dielectric constant measured in the transport and capacitive experiments.

\begin{figure}[htbp]
    \centering
    \includegraphics[width=0.1\textwidth]{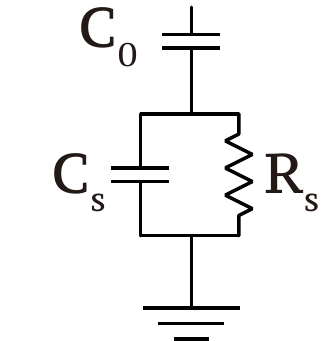}
    \caption{The lumped-element model of tip--sample interaction, with tip--sample capacitance $C_0$, sample capacitance $C_s$ and sample resistance $R_s$.}
    \label{fig:lumped}
\end{figure}

Next, we review the lumped-element model of tip--sample interaction. The sample is characterized by a capacitor $C_s$ in parallel with a resistor $R_s$. The tip is not in direct electrical contact with the sample, so an additional tip--sample capacitance $C_0$ is introduced between the tip and the sample (see Fig.~\ref{fig:lumped}) \cite{MIMReview}. It is straightforward to compute the admittance of this model,
\begin{align} \label{eq:lumpdef}
    Y_{\mathrm{lump}} & \sim \frac{1}{1 / j \omega C_{0}+R_{s} /\left(1+j \omega R_{s} C_{s}\right)}
\end{align}
where we use the electrical engineering convention with $j = -i$ rather than the quantum mechanical convention for time dependence. Now if we take 
\begin{equation}
    R_s = \frac{t}{A} \frac{1}{\Re \sigma_{\mathrm{eff}}}  = \frac{t}{A} \frac{1}{\omega \Im \epsilon_{\mathrm{eff}}} , \quad C_s =\frac{A}{t} \Re \epsilon_{\mathrm{eff}}
\end{equation}
where $A$ is the area of the sample, $\epsilon_{\mathrm{eff}} = \Re \epsilon_{\mathrm{eff}} - j \Im \epsilon_{\mathrm{eff}}$. Eqn.~(\ref{eq:lumpdef}) becomes
\begin{align} \label{eq:lump}
    Y_{\mathrm{lump}} & \sim \frac{j \omega}{1 / C_{0}+ \frac{j t/A \Im \epsilon_{\mathrm{eff}}}{1+j \Re \epsilon / \Im \epsilon_{\mathrm{eff}}}} \sim - i \omega C_0 \left ( 1 - \frac{t}{A} C_0 \frac{1}{\epsilon_{\mathrm{eff}}} \right )
\end{align}
in the limit $C_0 \ll \frac{A}{t} |\epsilon_{\mathrm{eff}}|$. If we compare Eqn.~(\ref{eq:diec}) and Eqn.~(\ref{eq:lump}), we find up to some prefactor and constant term, our framework agrees with the lumped-element model,
\begin{equation}
    Y_{\rchi} = Y_{\mathrm{lump}}
\end{equation}

\section{Density response function of the chiral edge mode} \label{sec:chichiral}

\subsection{Density response function of 1D chiral Luttinger liquid}

In this section, we first review the famous result of density response function of 1D chiral Luttinger liquid \cite{wen_theory_1992,wen_gapless_1991}, and present an alternative derivation following the spirit of Ref.~\onlinecite{shi_gifts_2022}. The density response function is first obtained by matching gauge invariance of the 1D boundary and its bulk \cite{wen_theory_1992},
\begin{equation}
    \rchi^0  (q) = \frac{1}{2 \pi} \frac{q}{\hbar \omega - \hbar v q}
\end{equation}
where $v$ is the Luttinger liquid velocity. This expression differs from the original form in Ref.~\onlinecite{wen_theory_1992} by a factor of $e^2$ due to a slightly different definition of $\rchi$.

Now we present an alternative derivation using constraints from anomaly. We first write down the action of a 1+1D chiral fermion,
\begin{equation} \label{eq:action}
    S= \int \dd t \dd x \psi^{\dagger} \big ( i \hbar (\partial_t + v \partial_x ) + e(A_0 + A_1) \big ) \psi
\end{equation}
which can be obtained by keeping only the left moving branch of a 1+1D masless Dirac fermion $S=\int \dd x^2 i \bar{\Psi} \slashed{\mathcal{D}} \Psi$. This action comes along with a $U(1)$ symmetry, $\psi \mapsto e^{i \theta} \psi$, which constrains the number current to be propotional to the number density, $ n = j = \psi^{\dagger} \psi$. This action also implies a chiral anomaly,
\begin{equation}
    \hbar \partial_t n + v \hbar \partial_x j = \frac{e E}{2\pi}
\end{equation}
Now we plug in $n = j$ and get
\begin{equation}
    - i \omega \hbar n + i v q \hbar n  = - \frac{e}{2\pi} i q A_0
\end{equation}
where we fix the gauge $\partial_t A_1 = 0$, then the density response function can be read off directly,
\begin{equation}
    \rchi^0(q) = \frac{1}{e} \frac{\partial n}{\partial A_0} = \frac{1}{2\pi} \frac{q}{\hbar \omega - \hbar v q}
\end{equation}
This result can be easily generalized to multiple branches,
\begin{equation}
    \rchi^0(q) = \sum_{\lambda} \frac{1}{2\pi} \frac{q}{\hbar \omega - \hbar v_{\lambda} q}
\end{equation}
where $v_{\lambda}$ is the velocity of each branch.

\subsection{Full response function of the chiral edge mode and its MIM response}

\label{sec:RPA}

The calculation above assumes a perfect 1D system. In reality, these edge modes will have a finite localization length $l$ into the bulk. For QAH insulators, this localization length $l$ is expect to inverse proportion to the bulk gap. However, even if we take the bulk gap to infinity, there will be a finite effective localization length $l$ since 1D system is not stable under long range Coulomb interaction. This can be seen by taking a naive 1D Fourier transform of the Coulomb interaction $1/4 \pi \epsilon_0 |\vect{r} -  \vect{r}'|$, where $G(\bq)$ diverges at all $\bq$. The presence of such localization length $l$ haven been well understood in theoretical works on the semiclassical correction to the edge theory \cite{VolkovEMP,Liang,MacDonaldEMP, Justin}.

We can incorporate this effect by computing the RPA version of the density response function \cite{MORINARI1996163},
\begin{equation}
    \rchi(q) = \frac{\rchi^0(q)}{1 - e^2 G(q) \rchi^0(q)}
\end{equation}
where $G(q) = -\frac{1}{2 \pi \epsilon_{0}} \log \left( q l \right)$ is the Fourier transformed  Coulomb interaction in 1D. Then the full density response function becomes
\begin{equation} \label{eq:QAHfinal}
    \rchi(q) = \frac{\frac{q}{2\pi}}{\hbar \omega - \hbar v q - \frac{e^2}{(2 \pi)^2 \epsilon_{0}} q \log \left( \frac{1}{q l} \right)}
\end{equation}
The poles of $\rchi(q)$ define the edge magnetoplasmon frequencies,
\begin{equation}
    \hbar \omega_{\mathrm{EMP}} =  \hbar v q + \frac{e^2}{(2 \pi)^2 \epsilon_{0}} q \log \left( \frac{1}{q l} \right)
\end{equation}
which agrees with the result in Ref.~\onlinecite{MacDonaldEMP} with $\hbar v = e E_e l^2$ in the context of quantum Hall systems. We can also rewrite $\rchi(q)$ in terms of $\hbar \omega_{\mathrm{EMP}}$,
\begin{equation} \label{eq:QAHEMP}
    \rchi(q) = \frac{1}{2\pi}\frac{q}{\hbar \omega - \hbar \omega_{\mathrm{EMP}}}
\end{equation}
For Chern bands with Chern number $C > 1$, we get
\begin{gather} \label{eq:C}
    \rchi(q) = \frac{1}{2\pi} \frac{C q}{\hbar \omega - \hbar v q - \frac{C e^2}{(2 \pi)^2 \epsilon_{0}} q \log \left( \frac{1}{q l} \right)}\\
    \hbar \omega_{\mathrm{EMP}} =  \hbar v q + \frac{C e^2}{(2 \pi)^2 \epsilon_{0}} q \log \left( \frac{1}{q l} \right)
\end{gather}
assuming that all branches have the same velocity and do not interact with each other. 

Now we are ready to compute the MIM response using the density response in Eqn.~(\ref{eq:QAHfinal}). We remind the readers that at microwave frequency, the EMP frequencies are actually discretized due to the quantization of momentum. We consider two different limits. One limit is when the MIM frequency $\omega$ is comparable to the first EMP frequency, then the dominant contribution just come from a particular momentum $k_{\omega}$ corresponding to the EMP frequency,
\begin{equation}
\begin{aligned}
    Y &\sim - \frac{i \omega e^2}P G(k_{\omega})^2 \rchi(k_{\omega}) \\
    &\sim - \frac{i \omega e^2}{32 \pi^3 \epsilon_0^2 P} \frac{ k_{\omega}}{\hbar \omega - \hbar \omega_{\mathrm{EMP}}(k_{\omega})} K_0^2(k_{\omega} r)\\
    &\sim - K_0^2(k_{\omega} r) \underset{r \to 0}{\sim} -\log^2 k_{\omega} r
\end{aligned}
\end{equation}
where in the last line we only keep the spatial $r$ dependence of the MIM response. The other limit is when the MIM frequency $\omega$ is much higher than the first few EMP frequencies, then we replace the summation over $k$ by an integral. To make analytical progress, we assume $\hbar \omega_{\mathrm{EMP}}(k) = \hbar v k$ is dominated by the quantum part,
\begin{equation}
\begin{aligned}
    Y &\sim - i \omega e^2 \int \frac{\dd k}{2 \pi} G(k)^2 \rchi(k) \\
    &\sim - \frac{i \omega^2 e^2}{256 \pi^{9/2} \epsilon_0 \hbar v^2} G_{3,5}^{5,2}\left(\frac{\omega ^2 r^2}{v^2}\bigg|
    \begin{matrix}
     -1/2,0,1/2 \\
     -1/2,0,0,0,0 \\
    \end{matrix}
    \right)\\
    &\sim -G_{3,5}^{5,2}\left(\frac{\omega ^2 r^2}{v^2}\bigg|
    \begin{matrix}
     -1/2,0,1/2 \\
     -1/2,0,0,0,0 \\
    \end{matrix}
    \right) \underset{r \to 0}{\sim} \frac1{r}
\end{aligned}
\end{equation}
where $G$ is the Meijer G function.

\section{Numerical simulation} \label{sec:simapp}

In this section we detail the numerical simulation setup. We implement the realistic device geometry, and solve the classical Maxwell's equation in frequency domain using the electric current module of the commercial COMSOL Multiphysics. The geometry we consider is shown in Fig.~\ref{fig:MIMSimu} (a). A tip with head diameter \SI{200}{nm} is suspended \SI{95}{nm} above the sample. The \SI{40}{nm} think \ce{Bi2Te3} sample lies on top of the silicon dioxide substrate. We add a back gate at the bottom of the substrate. We also extend the substrate \SI{4}{\um} beyond the sample, and include \SI{5}{\um} high of vacuum above the substrate to accommodate the long range Coulomb interaction between the tip and the sample. The dielectric constant of each region are set according to the material listed above. As explained in the main text and Ref.~\onlinecite{VolkovEMP,Justin}, the Hall conductance inside the \ce{Bi2Te3} sample is set up to $e^2/h$ to reproduce the density response of EMP modes. We also keep a small but finite longitudinal conductance $\sigma \sim \SI{3.8e-10}{S}$ inside the sample to avoid any divergence at EMP frequencies. This finite conductance in fact reflects the microscopic dissipation along the sample edge. In the main text, we always present the admittance $Y$ between the surface of the tip and the back gate.

One complication that prevents direct comparison between the simulation and the experiment at same frequency is the necessity to scale down the sample dimension to ensure numerical convergence. The complexity of this particular simulation is roughly determined by the number of mesh grids, which in turn is determined by the sample area divided by the sample thickness, which grows very quickly given that the sample is only \SI{40}{nm} thick. Therefore, we stick to a sample size of $\SI{12}{\um} \times \SI{20}{\um}$, and allows the mesh grid to vary between \SI{1}{nm} and \SI{30}{nm} inside the sample and between \SI{1}{nm} and \SI{15}{\um} outside. 
This rescaling of the sample size has a side effect of scaling up the first few EMP frequencies $\omega_n \sim \hbar \omega_{\mathrm{EMP}}(2 \pi n/P)$, which depends very sensitively on the sample perimeter (see Eqn.~(\ref{eq:EMP})).

There are two ingredients that are necessary to obtain EMP modes compared to usual setup used to simulate the MIM signal \cite{Eric}. The first is that the simulation has to be performed in a true 3D geometry instead of a 2D axisymmetric geometry since EMP modes explicitly breaks the rotation symmetry. The second and more subtly one is that a direct solver must be used to obtain EMP modes. This is because the EMP mode configuration is topologically nontrivial, and a conventional iterative solver cannot reach this regime and therefore end up converging to a local minimal in the topologically trivial regime.

\subsection{Effect of dissipation} \label{sec:dissipation}

\begin{figure}[htbp]
    \centering
    \includegraphics[width=0.3\textwidth]{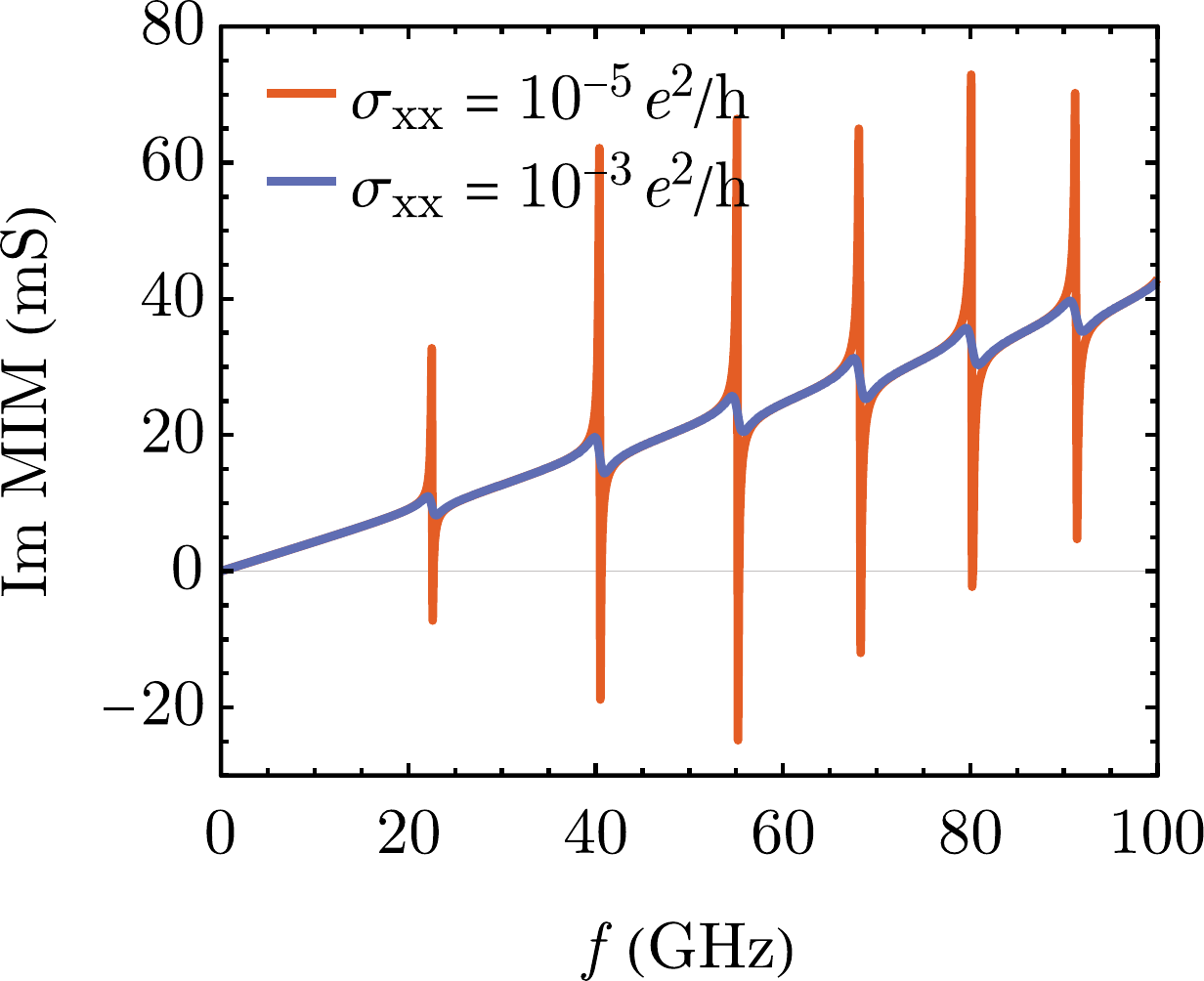}
    \caption{Imaginary part of the simulated MIM response with bulk longitudinal conductance $\sigma_{xx} = 10^{-5} e^2/h$ and $\sigma_{xx} = 10^{-3}e^2/h$, plotted as a function of frequency when the tip is positioned over the sample edge. The lineshape of the resonance peaks infers the level of dissipation along the sample edge.}
    \label{fig:dissipation}
\end{figure}

In this section, we explore the effect of dissipation on the MIM response in QAH insulators. One major source of dissipation presented along the sample edge originates from the bulk longitudinal conductance, which causes back scattering between counteracting edge modes at two sides of the sample. Then the density response function becomes
\begin{gather}
    \rchi(k) = \frac{1}{2\pi}  \frac{k}{\hbar \omega - \hbar v k - \frac{e^2}{(2 \pi)^2 \epsilon_{0}} k \log \left( \frac{1}{k l} \right) + i \epsilon}\\
    \Im Y_{\mathrm{ts}} \sim \Re \rchi(k_{\omega}) = \frac{k_{\omega}}{2\pi} \frac{\hbar \omega - \hbar \omega_{\mathrm{EMP}}(k_{\omega})}{(\hbar \omega - \hbar \omega_{\mathrm{EMP}}(k_{\omega}))^2 +  \epsilon^2}
\end{gather}
where $\epsilon \sim \sigma_{xx}$ characterizes the dissipation. We show this effect via the numerical simulation as shown in Fig.~\ref{fig:dissipation}. Considering the bulk conductance $\sigma_{xx} \sim 10^{-3}e^2/h$ of our sample, we expect the EMP resonances to be still clearly visible given the ability to sweep the MIM frequency continuously.

\subsection{EMP frequencies extracted from simulation} \label{sec:frequency}

\begin{figure}[htbp]
    \centering
    \includegraphics[width=0.22\textwidth]{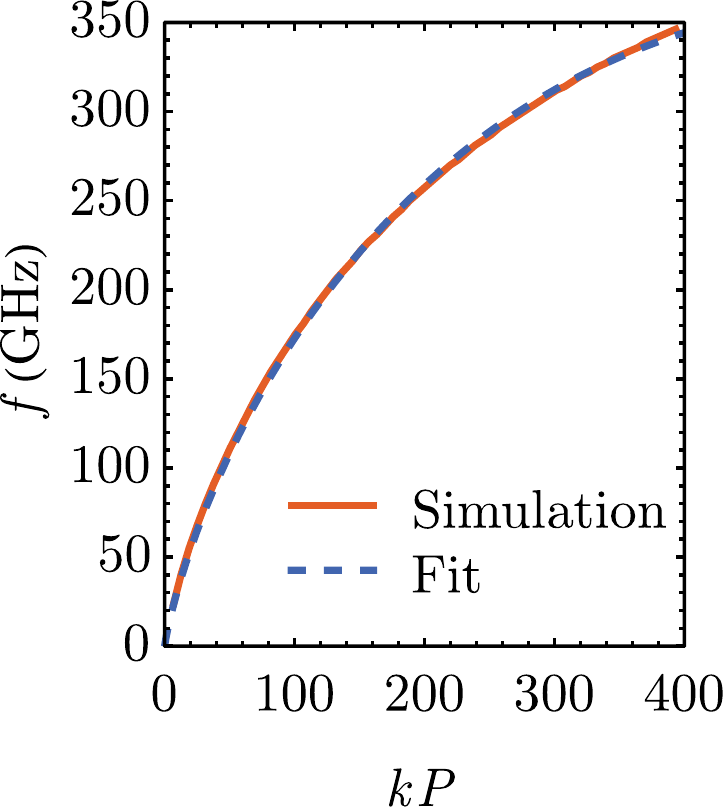}
    \caption{EMP frequencies extracted from the simulation. The fitting using Eqn.~(\ref{eq:EMP}) gives $\epsilon_r = 1.23$ and $l = \SI{54}{\nm}$.}
    \label{fig:frequency}
\end{figure}

In this section, we extract the EMP frequencies by identifying resonance peaks in $\Im Y_{\mathrm{ts}}$. As shown in Fig.~\ref{fig:frequency}, it fits well with the classical part of Eqn.~(\ref{eq:EMP}),
\begin{equation}
    \hbar \omega_{\mathrm{EMP}}^{\mathrm{class}} = \frac{e^2}{(2 \pi)^2 \epsilon_{0}} k \log \left( \frac{1}{k l} \right)
\end{equation}
since the formulation used in the simulation is semiclassical. Even though the Hall conductance is a step function across the edge, there is a finite edge localization length $l =\SI{54}{nm}$. This is due to the Coulomb instability explained in Sec.~\ref{sec:RPA} and is consistent with previous theoretical work on semiclassical corrections to the edge theory \cite{VolkovEMP,Liang,MacDonaldEMP, Justin}.

\section{MIM response of conductive edge defects} \label{sec:non-QAH}

\begin{figure}[htbp]
    \centering
    \includegraphics[width=0.48\textwidth]{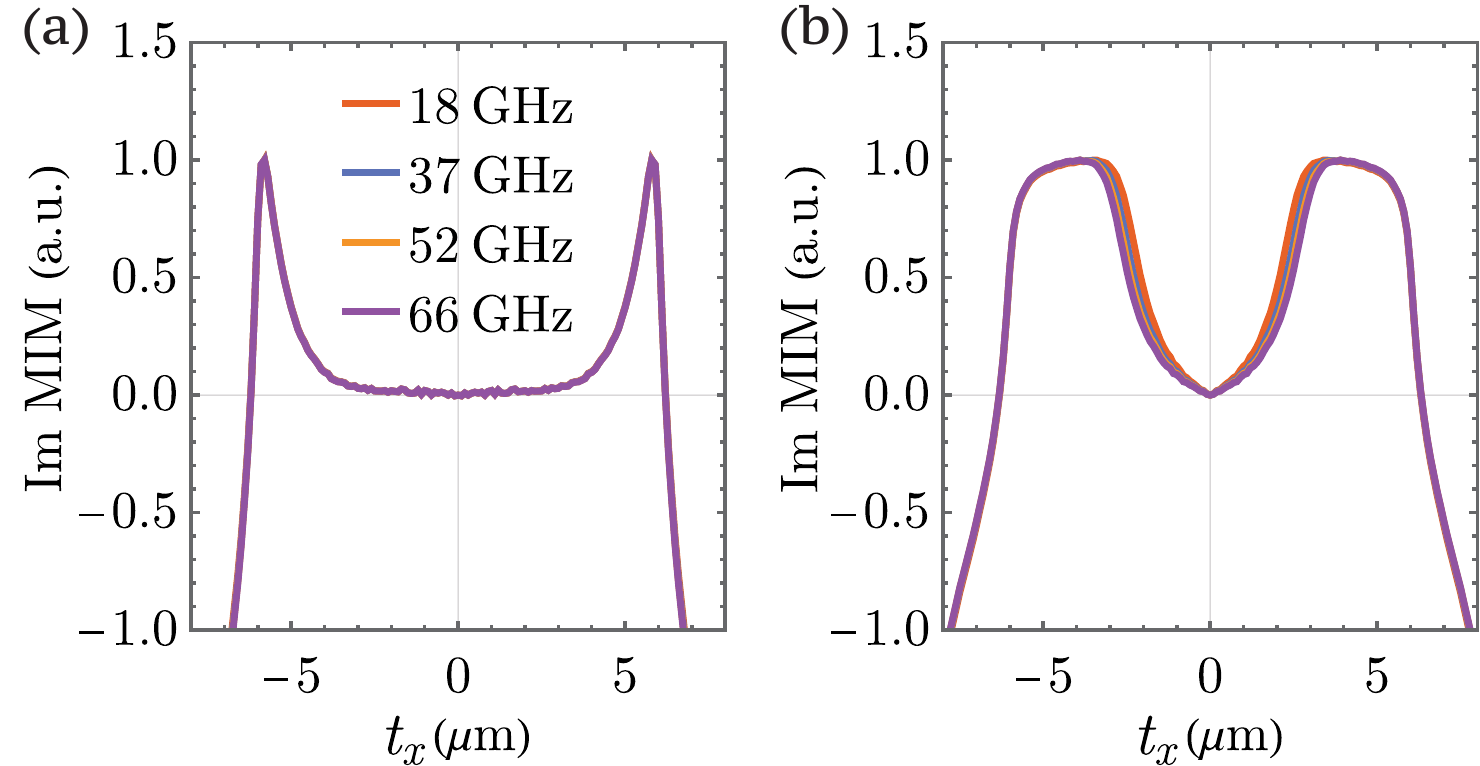}
    \caption{Simulated imaginary channel MIM signal across the sample with exponential decay conductivity (see text). The decay length $\xi$ in (a) is $\SI{10}{\nm}$ and the decay length in (b) is $\SI{100}{\nm}$.}
    \label{fig:1dmetal}
\end{figure}

\begin{figure*}[htbp]
    \centering
    \includegraphics[width=0.7\textwidth]{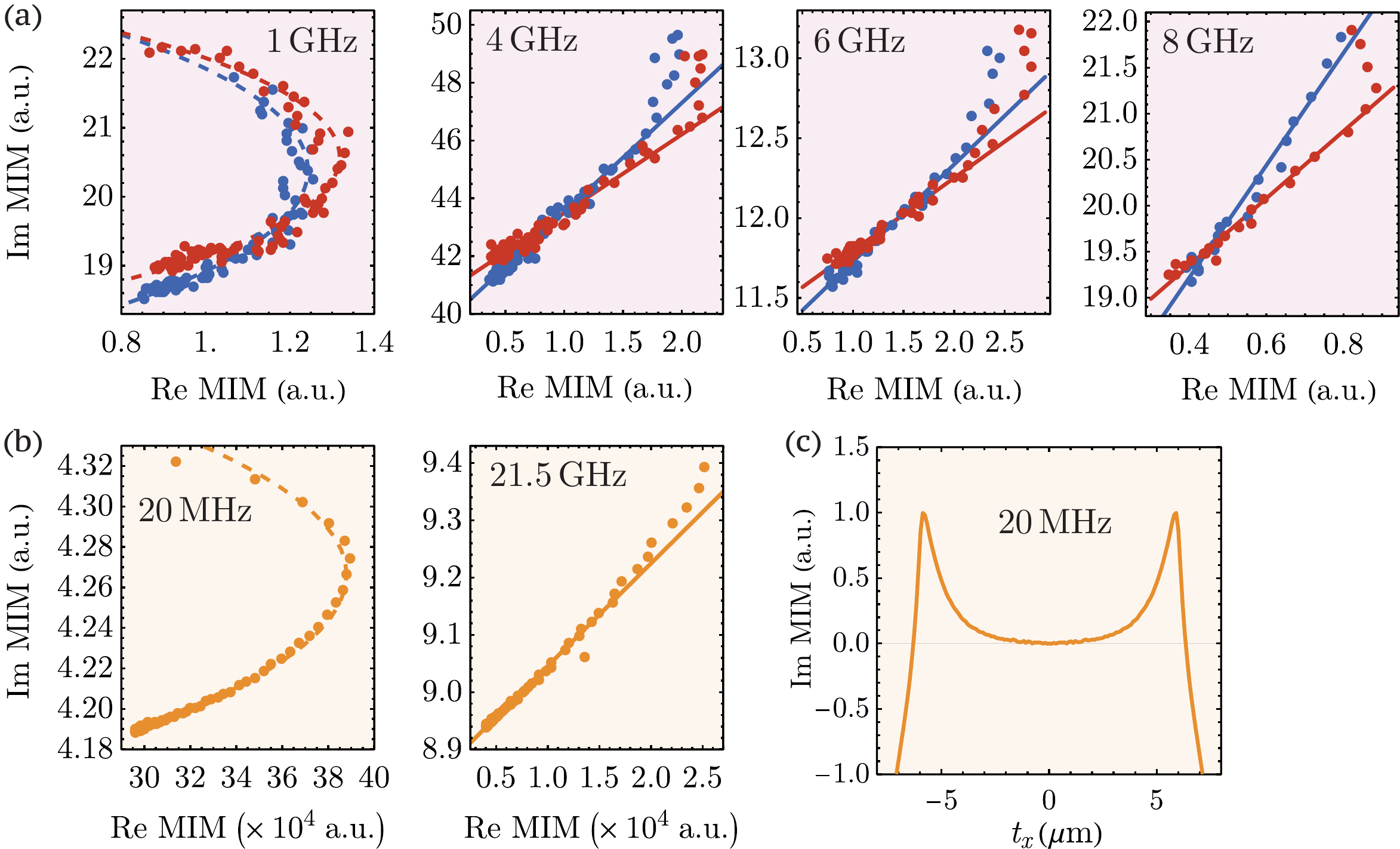}
    \caption{The $\Re \MIM - \Im \MIM$ scatter plot of (a) the experiment data at \SI{1}{\GHz}, \SI{4}{\GHz}, \SI{6}{\GHz}, and \SI{8}{\GHz} and (b) the simulation result way below the first EMP frequency (at \SI{20}{\MHz}) and around the first EMP frequency (at \SI{21.5}{\GHz}). The red (blue) curves correspond to data close to the left (right) sample edge. (c) Simulated imaginary channel MIM signal across the sample way below the first EMP frequency (at \SI{20}{\MHz}).}
    \label{fig:scatter}
\end{figure*}

To explore the difference in the MIM edge signal of topological edge modes and trivial impurity modes, we also numerically simulate the MIM signal of a trivial insulator with high conductance close to its boundary with identical setup described in Sec.~\ref{sec:simapp}. The conductance inside the trivial insulator takes the form
\begin{equation}
    \sigma(x) = \frac{\sigma_{\mathrm{tot}}}{\xi} e^{-x/\xi}
\end{equation}
where $x$ is the distance from the sample edge and $\xi$ is the decay length. The resulting imaginary channel MIM signal of two different $\xi$'s are presented in Fig.~\ref{fig:1dmetal}. In order to reproduce the a similar signal-to-noise ratio and spatial profile observed in the simulation and experiment of a QAH insulator, the decay length $\xi$ needs to be as small as \SI{10}{\nm} and the peak conductance $\sigma(x=0)$ needs to be as high as \SI{1e7}{S/m}. Given the fabrication procedure of our sample, this interpretation is highly unlikely.

\section{MIM scatter plot} \label{sec:scatter}

In this section, we discuss the relation between the imaginary channel and the real channel MIM signal in more details. In Fig.~\ref{fig:MIMExpt} (d), we present the scatter plot around the first few EMP frequencies, which exhibit characteristic straight line behavior consistent with the prediction that the MIM response comes from 1D edge modes. Now we turn our attention to a different limit when the MIM frequency $\omega \ll \omega_{\mathrm{EMP}}(2 \pi/P)$. In this case, contribution from the EMP modes will be suppressed by $1/\omega_{\mathrm{EMP}}(2 \pi/P)$ and then contribution from the dielectric background kicks in and starts to dominate.

This transition is indeed seen in the experiment. As shown in Fig.~\ref{fig:scatter} (a), when the MIM frequency is lowered to \SI{1}{GHz}, which is potentially lower than the first EMP frequency, the imaginary and real channel scatter plot reduces to a semi-circle as one would expect from the lumped-element model. To better understand this transition, we also show the scatter plot obtained from the numerical simulation in Fig.~\ref{fig:scatter} (b). Below the first EMP frequency, the scatter plot indeed has a semi-circle shape similar to the experiment plot at \SI{1}{\GHz}. We remind readers that the actual frequencies do not agree between the experiment and the simulation because the EMP frequencies are scaled up in the simulation to allow for a smaller sample dimension required in the simulation. We also point out that the edge peak is still visible in the MIM signal even when the scatter plot has been dominated by the background contribution (see Fig.~\ref{fig:scatter} (c)).

\subsection{Lumped-element model fit}

For completeness, we perform a lumped-element model fit of the scatter plot of the MIM data at \SI{1}{\GHz}. In the limit of fixed $R_s$, Eqn.~\ref{eq:lumpdef} reduces to the equation of a circle in the $\Re Y - \Im Y$ plane,
\begin{equation}
    (\Re Y - \Re Y_0)^2 + (\Im Y - \Im Y_0)^2 = r^2
\end{equation}
with $Y_0 = \ii \omega C_0\frac{C_0 +2C_s}{2(C_0 + C_s)}$ and $r = \frac{\omega C_0^2}{2(C_0+C_s)}$. To allow for $C_s$ to vary, we relax the constrain $\Re Y_0 = 0$ during the fitting.

For the fitting shown in Fig.~\ref{fig:scatter} (a) at \SI{1}{GHz}, we find $r = 0.046$ ($0.034$), $\Re Y_0 = -0.033$ ($-0.021$), and $\Im Y_0 = 0.203$ ($0.206$) for left (right) edge of the sample. At this level, negative $\Re Y_0$ is difficult to interprete within the lumped-element model.

\section{Experiment setup} \label{sec:divapp}

Following the procedure described in Ref.~\onlinecite{TokuraDis,Monica}, we use molecular beam epitaxy (MBE) to grow \ce{(Bi,Sb)2Te3} thin films with different \ce{Cr} doping within each layer on top of a semi-insulating \ce{InP}(111) substrate. In particular, the sample consists of a \SI{3}{nm} thick \ce{(Bi,Sb)2Te3} in between two layers of \SI{2}{nm} \ce{Cr-(Bi,Sb)2Te3}. After we remove the sample from the MBE chamber, we deposit a \SI{3}{\nm} thick \ce{AlO_x} capping layer by
the atomic layer deposition (ALD) system. We use photo-lithography and Ar-ion milling to isolate a \SI{400}{um} long and \SI{20}{um} wide Hall bar. Then we use chemical wet etching with an \ce{HCl-H3PO4} mixture to clean up the damaged side edge and substrate surface. The contact electrodes are fabricated by depositing a \SI{5}{nm} thick \ce{Ti} and subsequently a \SI{45}{nm} thick \ce{Au} by electron beam evaporation.

The MIM setup is identical to the one in Ref.
~\onlinecite{MIM2016,Monica}. We send microwaves at frequency $\SI{1056}{\MHz}$, $\SI{4144}{\MHz}$, $\SI{6327}{\MHz}$ and $\SI{8432}{\MHz}$ to the tip via an impedance matching network and collect the reflected signals. The impedance matching network is re-optimized every time when the frequency changes, which implies a different set of MIM constants $a$ and $b$ in general. The tip used in the experiment has diameter \SI{200}{\nm} and the tip--sample distance is maintained around \SI{95}{\nm}. The MIM data present in this work are obtained at \SI{450}{mK}.

\vfill

\end{document}